\documentclass[12pt]{article}
\usepackage{epsf}
\usepackage{amsfonts} 
\usepackage{amsbsy} 

\newdimen\hsgraph \newdimen\vsgraph
\hsgraph=\hsize \advance\hsgraph by-1.2truein
\vsgraph=\vsize \advance\vsgraph by-1.5truein

\newcommand{\C}[1]{{\mathcal #1}}

\newcommand{\R}[1]{{\mathrm #1}}

\newcommand{\beq}[1]{\begin{equation}{\label{#1}}}
\newcommand{\eeq}{\end{equation}}
\newcommand{\bea}{\begin{eqnarray}}
\newcommand{\eea}{\end{eqnarray}}

\newcommand{\nn}{\nonumber}
\newcommand{\rf}[1]{(\ref{#1})} 

\newcommand{\expect}[1]{{\left\langle #1\right\rangle}_{CE}}

\newcommand{\half}{{1\over 2}}
\newcommand{\threehalves}{{3\over 2}}
\newcommand{\fivehalves}{{5\over 2}}
\newcommand{\third}{{1\over 3}}
\newcommand{\fourthirds}{{4\over 3}}
\newcommand{\quarter}{{1\over 4}}
\newcommand{\eighth}{{1\over 8}}
\newcommand{\twothirds}{{2\over 3}}

\newcommand{\gstr}{\gamma_{str}}

\newcommand{\new}{}
\newcommand{\ts}[1]{{\textstyle #1}}
\newcommand{\scs}[1]{{\scriptstyle #1}}
\newcommand{\BP}{$\C B$}
\newcommand{\Pb}{{\bar P}}
\newcommand{\Qt}{{\widetilde{\C Q}}}
\newcommand{\Pt}{{\widetilde{\Phi}}}
\newcommand{\RB}{{\R B}}
\newcommand{\RA}{{\R A}}
\newcommand{\RC}{{\R C}}
\newcommand{\diff}{{\left(\frac{d~}{dy}\right)}}
\newcommand{\yy}{{\Bigg |_{y=1}}}
\newcommand{\BtoC}{{({\scriptstyle \RB\to \RC})}}
\newcommand{\sumprime}{\mathop{{\sum}'}}

\begin{document}
\topmargin 0pt
\oddsidemargin 5mm
\headheight 0pt
\topskip 0mm

\addtolength{\baselineskip}{0.20\baselineskip}

\pagestyle{empty}

\begin{flushright}
OUTP-97-33P\\
October 1997\\
hep-th/9710024
\end{flushright}

\begin{center}

\vspace{18pt}
{\Large \bf The  Spectral Dimension of the Branched Polymer Phase of
Two-dimensional Quantum Gravity}

\vspace{2 truecm}

{\sc Thordur Jonsson\footnote{e-mail: thjons@raunvis.hi.is} }\\
{\em Raunvisindastofnun Haskolans, University of Iceland \\
Dunhaga 3, 107 Reykjavik \\
Iceland\\}
\vspace{1 truecm}
{\sc John F. Wheater\footnote{e-mail: j.wheater1@physics.ox.ac.uk}}


{\em Department of Physics, University of Oxford \\
Theoretical Physics,\\
1 Keble Road,\\
 Oxford OX1 3NP, UK\\}

\vspace{2 truecm}

\end{center}

\noindent
{\bf Abstract.} The metric of two-dimensional  quantum gravity 
interacting with 
conformal matter is believed to collapse to
 a branched polymer metric when the central
charge $c>1$. We show analytically  that the spectral dimension, $d_S$,
 of such  a branched polymer
phase is $\fourthirds$. This is in good agreement with numerical simulations
for large $c$.

\vfill
\begin{flushleft}
\end{flushleft}
\newpage
\setcounter{page}{1}
\pagestyle{plain}
\section{Introduction}
The nature of the dimensionality of the manifolds appearing in euclidean
quantum gravity ensembles has attracted considerable attention recently
\cite{Kawai,AmbA,Catt,Bow,AmbB,AmbC,AmbE}. There 
are, in principle, many ways of defining the dimension; on smooth regular
manifolds we expect every definition to yield the same value (for example
2 if the system is ``two-dimensional''). However the euclidean
quantum gravity ensembles contain many members which are very far from smooth,
often showing some kind of fractal structure, and different definitions of
dimension probe different global aspects of the geometry. Thus they
 can (and sometimes  do) yield different numerical  values.

Although these questions are of very general applicability by far the most 
complete studies of the notion of dimension have been conducted for 
two-dimensional euclidean quantum gravity (here ``two-dimensional'' means
that every manifold in the ensemble is of dimension two in terms of
the standard mathematical definition of a manifold).  The ensemble is
defined at fixed volume (which we denote the canonical ensemble, CE for short)
by the partition function
\beq{i1} Z(V;\lambda)=\int Dg\, \delta\left(\int d^2\xi\,{\sqrt g}-V\right)
\exp\left(-S_{eff}(g;\lambda)\right)\eeq
where the functional integral runs over all physically inequivalent 
metrics $g$ with the volume constrained by the delta-function to be $V$. 
The functional $S_{eff}(g;\lambda)$ is the effective action obtained for the metric after 
integrating out all
matter fields 
 and $\lambda$ represents any parameters (for example the central charge $c$)
 describing the
matter field theory.  Expectation values are given by
\beq{i2}
\expect{\cdot}=\frac{1}{Z(V;\lambda)}
\int Dg\,(\cdot)\, \delta\left(\int d^2\xi\,{\sqrt g}-V\right)
\exp\left(-S_{eff}(g;\lambda)\right)\eeq
with the understanding that it only makes sense to calculate the expectation
values of reparametrization invariant quantities.  We can also define a grand
canonical ensemble (or GCE) partition function
\beq{i3}\C Z(\mu;\lambda)=\int_0^\infty e^{-\mu V}Z(V;\lambda)\,dV.\eeq
In practice nearly all the available results about the dimensionality of
the manifolds in these ensembles have been obtained in the 
discretized formulation \cite{AmbD,Kazakov,David}.
  The functional integral over metrics of a 
given volume is replaced by the sum over all 
triangulations (or, more generally, graphs), $\C G$, with the number of 
 vertices, $N_{\C G}$, fixed so that the CE partition function is
\beq{x1}  Z(N;\lambda)=\sum_{{\C G}}\delta_{N,N_{\C G}}
\exp\left(-S_{eff}({\C G};\lambda)\right)\eeq
and expectation values given by
\beq{x2}
\expect{\cdot}=\frac{1}{Z(N;\lambda)}
\sum_{{\C G}}\delta_{N,N_{\C G}}(\cdot)
\exp\left(-S_{eff}({\C G};\lambda)\right).\eeq
To calculate the GCE partition function
 the integral over $V$ in \rf{i3} is 
replaced by a sum
over $N$.

Two different notions of dimension have been considered in detail.  The first
is the ``Hausdorff dimension'', $d_h$, which is defined by considering the 
volume of space, $d\tau$, contained in a shell of geodesic radius $R$ and
thickness $dR$; provided that $R$ is much bigger than the short distance
cut-off and much smaller than the characteristic linear geodesic size
(which is $V^{1/\nu}$ for some exponent $\nu$) we expect that
\beq{i4}\expect{d\tau}\sim R^{d_h-1}dR.\eeq
For $c=0$ an analytic  calculation has been done and it is found that
$d_h=4$ \cite{AmbA}.  For other central charges $c>0$  the situation is still 
unclear but numerical results indicate that $d_h=4$ for unitary $c<1$ 
theories \cite{AmbB} and there is some analytic support for this \cite{Bow}.
For $c>1$ it is believed that the ensemble of universes collapses to a
branched-polymer (BP) like phase; for a pure BP ensemble analytic
calculation shows that $d_h=2$ \cite{BP2}
  which is in very good agreement with 
numerical simulation for $c=5$ \cite{AmbE} (which seems to be the value of $c$ above which all these results change very little).

The second notion of dimension which has been investigated is the ``spectral dimension'', $d_S$. This is defined for a given smooth metric
 $g$ on a d-dimensional manifold by the 
behaviour of the
coincidence limit of the heat kernel $K_g(t;\xi,\xi')
$ for which it is known that
\beq{i5} K_g(t;\xi,\xi)=\frac{1}{t^{d/2}}\sum_{r=0}^\infty t^r a_r(\xi)\eeq
where the functions $a_r(\xi)$ are reparametrization invariant, see e.g. \cite{dewitt}.
Thus in the quantum gravity ensemble we can define $d_S$ by
\beq{i6}\expect{ K_g(t;\xi,\xi)}\sim\frac{1}{t^{d_S/2}}\eeq
for $t$ small enough that the random walks generated by the
Laplacian
do not probe the finite size structure of the manifolds.
In the discretized formulation we consider a random walk
on a graph $\C G$ with $N$ sites \cite{AmbE}; the random walker
moves from one lattice point to one of its neighbours at each time step with
equal  probabilities corresponding to diffusion governed by the Laplacian of the graph.
Letting the probability that the walker
 has returned to the starting point $i$ 
after $t$ steps be $P_{\C G}(i;t)$ we can define $d_S$ by
\beq{i7}P_N(t)=\expect {N^{-1}\sum_{i=1}^{N} P_{\C G}
(i;t)}\sim \frac{1}{t^{d_S/2}}.\eeq
This behaviour is expected to hold provided
 $0\ll t\ll N^{1/\Delta}$ for some exponent $\Delta$ so that discretization
and finite size effects are avoided. Numerical simulations show that 
$d_S\approx 2$ for
$c=0$  and that as $c$ increases $d_S$ decreases \cite{AmbE}. 
It was 
conjectured in \cite{AmbE} that $d_h=2d_S$ for all values of $c$.
  This would
imply that if the large $c$ phase is a BP phase  then it has $d_S=1$.
The purpose of this paper is to compute $d_S$ analytically for a pure branched
polymer phase in order to check this conjecture and provide some
definite result with which to compare the numerical simulations.

This paper is organized as follows. In section 2 we explain how $d_S$ which 
is defined in the CE can be calculated in the GCE. In section 3 we define
the branched polymer ensemble and find relations satisfied by
the  return probabilities. In section 4 we do the GCE sum and show that 
 $d_S=\fourthirds$.
Section 5 contains the proofs of various results used in section 4; some technicalities are relegated to the appendix.
Finally in section 6 we discuss our result and its connection to other
spectral dimension problems (in particular on percolation clusters) that 
have been considered in the past.

\section{Spectral dimension in the CE and the GCE}
Our calculation of the spectral dimension for the BP ensemble (which we will
define carefully in section 3) depends crucially upon the relationship between
the CE and the GCE. This is most easily understood in the following way
 (we assume
from now on that we are working with the discretized formulation).
Consider first the CE; at very large times $t\gg N^{\Delta}$, where $\Delta$
is a suitable exponent, the walk will
have  uniform probability of being at any of the $N$ lattice sites
(essentially this is just the contribution of the zero mode of the Laplacian) so
\beq{ii1} \lim_{t\to\infty} P_N(t)=\frac{1}{N}.\eeq
At much smaller times we expect the behaviour \rf{i7} so that
\beq{ii1a}\lim_{N\to\infty} P_N(t)\sim\frac{1}{t^{d_S/2}}
\eeq
A suitable interpolating
function between these two limits is
\beq{ii2} P_N(t)=\frac{1}{N}+ \frac{a}{t^{d_S/2}}
\exp\left(-\frac{t}{N^{\Delta}}\right)\eeq
where $a$ is a constant.  From this we can compute the generating function
\beq{ii3} \Pb_N(y)=\sum_{t=0}^\infty y^tP_N(t)\approx \frac{1}{N}\frac{1}{1-y}
+\frac{a}{(1-y+N^{-\Delta})^{1-d_S/2}}\eeq
where we have assumed that $d_S< 2$ and only written the dominating terms in
$\Pb_N(y)$ as $y\uparrow 1$ and $N\to\infty$.
Note the appearance of the simple pole at $y=1$ as a consequence of the zero 
mode and that the rest of $\Pb_N(y)$ is analytic at $y=1$ for finite $N$.
Defining $\Pb\,'_N(y)$ to be $\Pb_N(y)$ with 
 the contribution of the
zero mode subtracted  (which we will find is straightforward in our actual
calculation)  we see that
\beq{ii4} \lim_{N\to\infty}\Pb\,'_N(y)=\frac{a}{(1-y)^{1-d_S/2}}\eeq
so that, as we might expect, the non-analyticity associated with $d_S$ appears
only in the thermodynamic limit. Note that we also have that at large $N$
\beq{ii4a} \Pb\,'_N(1)\sim N^{\Delta(1-d_S/2)}.\eeq
This result depends only on there being a cut-off in 
 the $t^{-d_S/2}$ behaviour at  $t\sim N^\Delta$ and not on its
 particular form.

 In the GCE the natural quantity to compute
is
\beq{ii5}\C P(z,y)=\sum_{\C G} z^{N_{\C G}}\sum_{t=0}^\infty y^tP_{\C G}(i;t)\eeq
where the sum runs over all graphs $\C G$ in the ensemble, $N_{\C G}$ is the 
number of points in $\C G$ and, as before,  $P_{\C G}(i;t)$ is the return 
probability of a walker on $\C G$ returning to site $i$ after $t$ steps 
(in general
one would also sum over starting points $i$ but in the case of branched 
polymers it is sufficient to consider walks starting at the root and 
summation over $i$ is not necessary). The function $\C P$ is related to 
the CE quantity $\Pb_N(y)$ through
\beq{ii6} \C P(z,y)=\sum_{ N} z^{N}Z(N;\lambda)\Pb_N(y).  \eeq
We expect that for large $N$ the CE partition function behaves as
\beq{ii7} Z(N;\lambda)\sim N^{\gamma_{str}-2} z_0^{-N}\eeq
where $z_0$ is a constant and $\gamma_{str}\le\half$.  It follows that the 
coefficient of the pole term is finite at criticality (ie at $z\uparrow z_0$
 where graphs of arbitrary size contribute).  Dropping the pole term to get 
\beq{x4} \C P'(z,y) =\sum_{ N} z^{N}Z(N;\lambda)\Pb\,'_N(y)  \eeq
and approximating the sum over $N$ by an integral we find, using \rf{ii3},
\beq{ii8} \C P'(z,y)\sim\left(1-\frac{z}{z_0}\right)^\beta
\Phi\left(\frac{1-y}{\left(1-\frac{z}{z_0}\right)^\Delta}\right)\eeq
where the prefactor exponent $\beta$ is given by
\beq{ii9} \beta=1-\gamma_{str}+\Delta\left(\frac{d_S}{2}-1\right)\eeq
and the function $\Phi$ by
\beq{x5}\Phi(v)=\int_1^\infty\frac{a e^{-x}}{(v+x^\Delta)^{1-d_S/2}}\, dx.\eeq
Note that $\Phi(v)$ is analytic for $v\ge 0$ with $\Phi(0)=O(1)$. The result
\rf{ii8} describes correctly
 the singularity of $\C P'(z,y)$ at $z=z_0$ 
 provided that the prefactor diverges as $z\uparrow z_0$;
if it does not then we consider instead a higher derivative of 
$\C P'(z,y)$ with respect to $z$.  We draw three lessons from \rf{ii8}:
\begin{enumerate}
\item It is not the case that 
\beq{ii10}  \C P'(z,y)\stackrel{ y\to 1}{\sim}
\frac{1}{(1-y)^{1-d_S/2}}\eeq
In fact, for $z<z_0$, $\C P'(z,y)$ is an analytic function on
a neighbourhood of $y=1$.
\item The exponent $\Delta$ is a \emph{gap exponent}; that is the $n+1$st
derivative of $\C P'(z,y)$ with respect to $y$ is more divergent
than the  $n$th derivative as $z\uparrow z_0$ by a factor
$(1-z/z_0)^{-\Delta}$.
\item If we can compute the prefactor exponent, $\beta$, and the gap exponent
$\Delta$ in the GCE we can find $d_S$ using \rf{ii9}.
\end{enumerate}
\noindent This discussion is based on the interpolating ansatz \rf{ii2}
and the reader might worry that it is not completely general. In fact,
although the ansatz is useful for pedagogical purposes, we shall show 
in section 4.2 that it is not necessary.

\section{Branched Polymers}
\subsection{The Polymer Ensemble}
In this work we will restrict our attention to polymers which have
 vertices of order one or of order three and in which one 
vertex of order one, 
called the root, is distinguished.
This is the simplest case of the
generic rooted branched polymer ensemble \cite{BP2}. The simplest polymer, $\R B_0$,
consists of one link joining the root, labelled  ``0'', to one other vertex
 of order one which we label ``1'' as shown in fig.1.
\begin{figure}[h]
{\epsfxsize=\textwidth \epsfbox{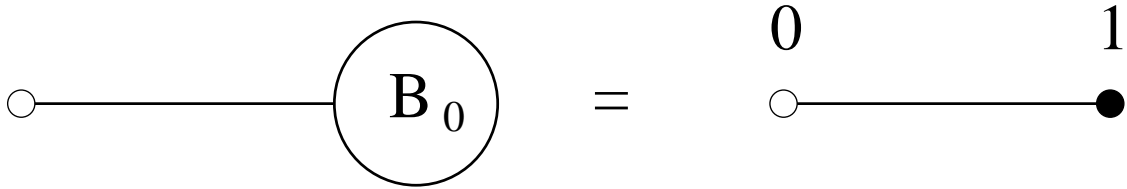}}
\caption{The elementary polymer $\RB_0$.}
\end{figure}
\noindent The ensemble of all polymers, \BP, is generated by the statements:
\begin{enumerate}
\item The only element of {\BP} consisting of one link is $\R B_0$.
\item Any two polymers,   $\R B_1$
 and $\R B_2$ may be combined to generate a new, and larger, polymer
 $\R B_3=\R B_1\cup\R B_2$ as shown in fig.2.
\end{enumerate}
\begin{figure}[h]
{\epsfxsize=\textwidth \epsfbox{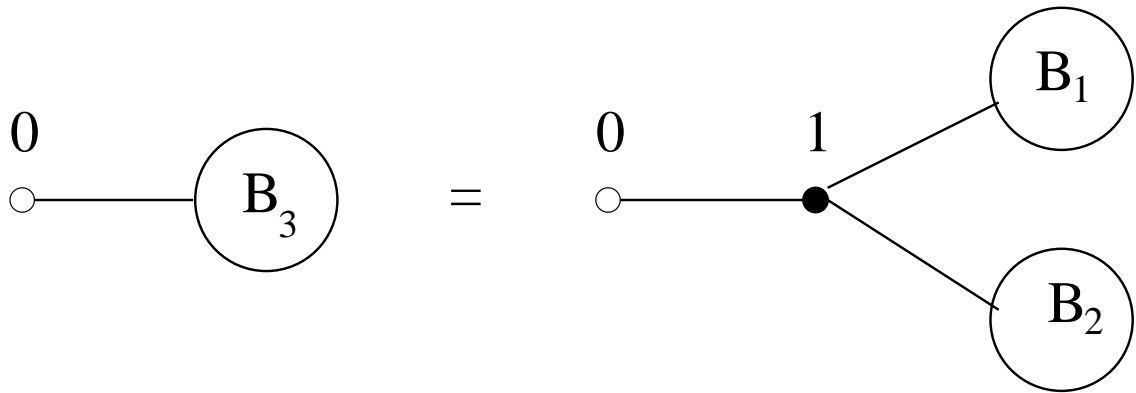}}
\caption{The combination of polymers $\RB_1$ and $\RB_2$ to form $\RB_3$.}
\end{figure}
It is clear that any polymer $\R B\in\C B$, ${\RB}\ne {\RB}_0$, has a unique decomposition into
two constituent polymers $\R B'$ and $\R B''$ by cutting just to the right of vertex ``1'' as shown in fig.3.
\begin{figure}[h]
{\epsfxsize=\textwidth \epsfbox{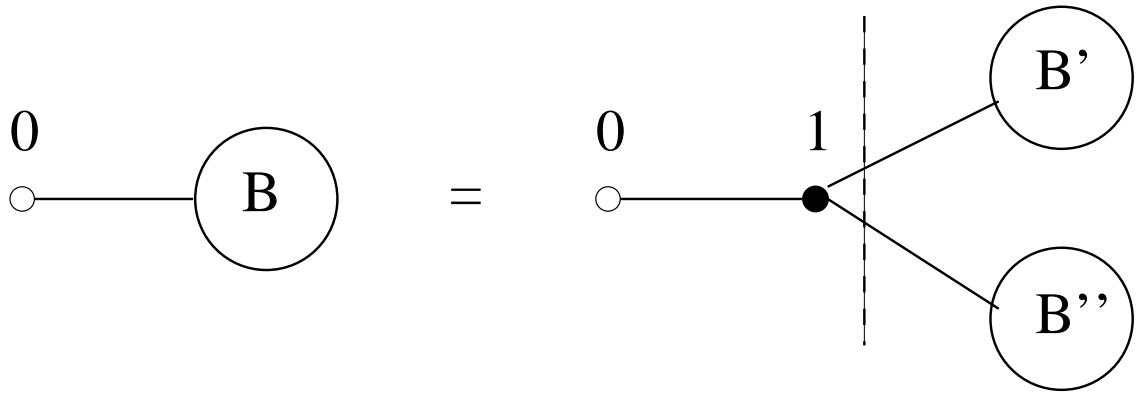}}
\caption{The unique decomposition of polymer $\RB$ to its constituents
$\RB'$ and $\RB''$.}
\end{figure}

One characteristic of an element $\R B$ of {\BP} is the number of order one 
vertices
(not counting the root) which we will call the \emph{size} and 
denote $N_B$; clearly  we have
\beq{2.1} 
N_{\R B_1\cup\R B_2}=N_{\R B_1}+N_{\R B_2}.
\eeq
Letting $\Omega_N$ be  the number of elements of {\BP} with a given size, $N$,
we find
\beq{2.2}
\Omega_N=\sum_{M=1}^{N-1}\Omega_M\Omega_{N-M}
\eeq
with $\Omega_1=1$. This is the recurrence relation for the Catalan numbers,
$T_N$, so
\beq{3.3}
\Omega_N=T_N=\frac{(2N-2)!}{N! (N-1)!}.
\eeq
For large $N$, the numbers $T_N$ have the asymptotic behaviour
\beq{3.4}
T_N\sim\frac{4^N}{N^{\frac{3}{2}}}.
\eeq
Note that the number of vertices of order three is $N-1$, and hence the total
number of vertices excluding the root is $2N-1$, so $N$ is a sensible measure
of the volume of the system. We therefore expect on general grounds that
$\Omega_N\sim N^{\gstr-2} \exp\mu_0 N$; comparing with \rf{3.4} we recover the
well known result for branched polymers that 
$\gstr=\half$. The generating function of the  Catalan numbers will appear repeatedly in the rest of
this paper and is given by
\beq{3.5}
\C T(z)=\sum_{N=1}^\infty z^N T_N=\half(1-\sqrt{1-4z}).
\eeq
Note that this definition of {\BP} naturally
connects ensemble members of different sizes suggesting that
the Grand Canonical Ensemble (GCE), which contains polymers of all sizes,
may be easier to calculate with than the Canonical Ensemble (CE) which 
contains polymers of fixed size; indeed \rf{3.5} is nothing but the GCE 
partition function for {\BP} with the action $N_{\RB}$.

 \subsection{Random Walks on Polymers}

Since $\C B$ is defined with  a vertex of special status, the root,
it is convenient to consider walks starting and ending at the root rather
than summing over all vertices. This does not affect the value
of $d_S$ because in fact half of all vertices have coordination number one
like the root and are therefore equivalent as 
starting points.
First consider an arbitrary polymer ${\RB}\ne {\RB}_0$.
 At $t=0$ a walker sets out from the root;
at $t=1$, after taking one step, he/she is necessarily at the first vertex; 
for the next step  one of the three links attached to the first vertex is 
chosen
with uniform probability and the walk proceeds down that link. The process
continues in this way with the step taken chosen with uniform probability
from the possible steps. Now let $P_\RB^1(t)$ be the probability that the walker 
returns to the root \emph{for the first time} after $t$ steps, and let
$P_{\RB}(t)$ be the probability that, after $t$ steps, the walker is at the root.
Note that both $P_\RB^1(t)$ and   $P_\RB(t)$ are zero for odd $t$  and that 
$P_\RB^1(0)=0$ because at least two steps are needed to return to the root
for the first time.  Clearly walks contributing to $P_\RB(t)$ may be back at the
origin for the first, second, third .... occasion so we have
\beq{3.6}
P_\RB(t)=\delta_{t,0}+P_\RB^1(t)+\sum_{t_1+t_2=t}P_\RB^1(t_1)P_\RB^1(t_2)
+\ldots+\sum_{t_1+t_2+\ldots+t_n=t}\prod_{i=1}^n P_\RB^1(t_i)+\ldots\,.
\eeq
Introduce the generating functions
\bea
\Pb_\RB(y)=\sum_{t=0}^\infty y^{t/2}P_\RB(t)\label{3.7}\eea
and
\bea\Pb_\RB^1(y)=\sum_{t=0}^\infty y^{t/2}P_\RB^1(t).\label{3.8}
\eea
We substitute \rf{3.6} into \rf{3.7} and then use \rf{3.8} to obtain
\bea
\Pb_\RB(y)&=&1+\Pb_\RB^1(y)+(\Pb_\RB^1(y))^2+\ldots\nn\\
&=&\frac{1}{1-\Pb_\RB^1(y)}.\label{3.9}
\eea

Let us now consider a polymer ${\R A}\ne {\R B}_0$ which is made by joining two polymers
B and C, cf. fig.2.
We want to relate the first return probability  on A  to the first return
probabilities on its constituents B and C. A typical walk leaves the root
and then goes out and back many times from the first vertex along B and C
before finally returning to the root; an example is shown in fig.4.
\begin{figure}[h]
{\epsfxsize=\textwidth \epsfbox{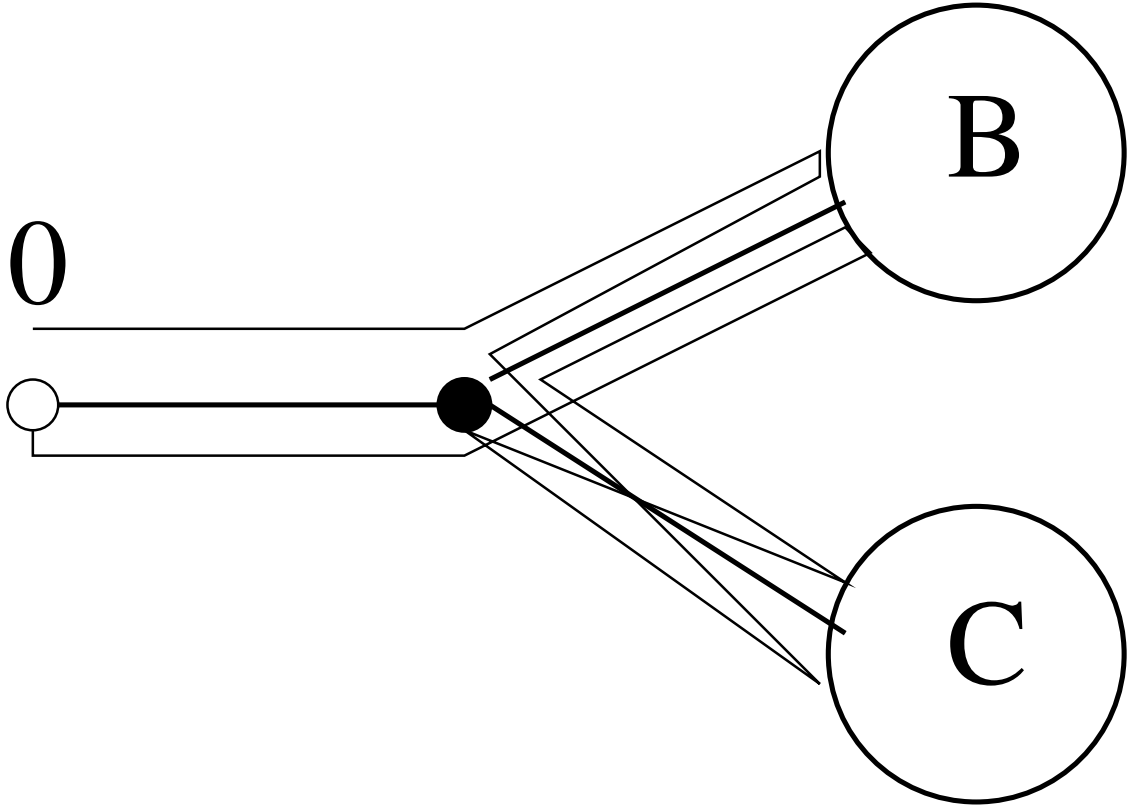}}
\caption{Example of a walk contributing to the first return probability
on polymer $\RA=\RB\cup\RC$.}
\end{figure}
Taking all these walks into account we find that
\bea
P_\RA^1(t)&=&\third\delta_{t,2}+\left(\third\right)^2\left(P_\RB^1(t-2)+P_\RC^1(t-2)
\right)\nn\\
&&+\left(\third\right)^3\sum_{t_1+t_2=t-2}\left(P_\RB^1(t_1)+P_\RC^1(t_1)
\right)\left(P_\RB^1(t_2)+P_\RC^1(t_2)
\right)+\ldots
\nn\\
&&\ldots+\left(\third\right)^{n+1}\sum_{t_1+\ldots+t_n=t-2}
\prod_{i=1}^n\left(P_\RB^1(t_i)+P_\RC^1(t_i)
\right)+\ldots\,.\label{3.10}
\eea
Inserting \rf{3.10} into \rf{3.8} we get
\bea
\Pb_\RA^1(y)&=&\third y+\left(\third\right)^2\left(\Pb_\RB^1(y)+\Pb_\RC^1(y)
\right)+\left(\third\right)^3\left(\Pb_\RB^1(y)+\Pb_\RC^1(y)
\right)^2+\ldots
\nn\\
&&\ldots+\left(\third\right)^{n+1}\left(\Pb_\RB^1(y)+\Pb_\RC^1(y)
\right)^n+\ldots\nn\\
&=&\frac{y}{3-(\Pb_\RB^1(y)+\Pb_\RC^1(y))}.\label{3.11}
\eea
It is convenient to introduce a new quantity $h_\RA(y)$ defined by
\beq{3.12}
h_\RA(y)=\frac{1}{1-y}\left(1-\Pb^1_\RA(y)\right).
\eeq
We find from \rf{3.9} that the return probability generating function
 on A is then given by 
\beq{3.13}
\Pb_\RA(y)=\frac{1}{1-y}\,\frac{1}{h_\RA(y)}
\eeq
and the recurrence relation for the first return probabilities \rf{3.11}
becomes
\beq{3.14}
h_\RA(y)=\frac{1+h_\RB(y)+h_\RC(y)}{1+(1-y)\left(h_\RB(y)+h_\RC(y)\right)}.
\eeq
Note that for the elementary polymer, $\R B_0$, we have
\beq{3.15}
h_{\R B_0}=1.
\eeq
We can, at least in principle, generate $\Pb_\RB(y)$ for every $\R B\in \C B$
from the recurrence \rf{3.14} and the initial condition \rf{3.15}.

\section{Random Walks in the GCE}
\subsection{The generating function}

We define the GCE generating function
\beq{3.16}
\C Q(z,y)=\sum_{\R B\in \C B} z^{N_{\R B}}\Pb_{\R B}(y)
\eeq
which, because of \rf{3.13}, becomes
\beq{3.17}
\C Q(z,y)=\frac{1}{1-y}\sum_{\R B\in \C B} z^{N_{\R B}}\left(h_{\R B}(y)\right)^{-1}.
\eeq

First we will show that $\C Q(z,y)$ has a simple pole at $y=1$ with finite
residue provided $z\le\quarter$. From \rf{3.14} we see easily by induction
that the functions $h_{\RB}$ are non-decreasing. It therefore
suffices to show that the sum in
\rf{3.17} with $h_{\R B}(y)$ replaced by $h_{\R B}(1)$ converges.  From 
\rf{3.14} we have
\beq{3.18}
h_{\R B\cup\R C}(1)=1+h_{\R B}(1)+h_{\R C}(1).
\eeq
From this, the initial condition \rf{3.15}, and the relation between
the sizes \rf{2.1}, it follows that
\beq{3.19}
h_{\R A}(1)=2N_{\R A}-1.
\eeq
Note that this quantity takes the same value for all polymers of a given
size. Using \rf{3.19} we have
\beq{3.20}
(1-y)\C Q(z,y)\vert_{y=1}=\sum_{N=1}^\infty \frac {z^N\Omega_N}{2N-1}
\eeq
which converges for $z\le\quarter$ on account of \rf{3.4}. 

As we discussed in section 2, the simple pole in $\C Q(z,y)$ at $y=1$ arises
from walks that are much longer than the size of the polymer; we expect to
find  these
walks at an arbitrary point in the polymer with essentially uniform 
probability which in this case, according to  \rf{3.19}, is
$(2N_{\R A}-1)^{-1}$.  The information about $d_S$ is contained in the
remaining (ie non-pole) part of $\C Q(z,y)$ so we define a new function
\bea
\Qt(z,y)&=&-\frac{d~}{dy}\left((1-y)\C Q(z,y)\right)\nn\\
&=&\sum_{\R B\in \C B} z^{N_{\R B}}\left(h_{\R B}(y)\right)^{-2}
\frac{dh_{\R B}(y)}{dy}.\label{3.21}
\eea
Now define
\bea\Qt_n(z)&=&\diff^{n}\Qt(z,y)\yy\nn\\
&=&-\sum_{\RB\in\C B} z^{N_\RB}\diff^{n+1}\frac{1}{h_\RB(y)}\yy.\label{n1}
\eea
Assuming for the moment that all these objects actually exist, then
formally
\beq{n2}\Qt(z,y)=\sum_{n=0}^\infty\frac{(y-1)^n}{n!}\Qt_n(z).\eeq
We show in the appendix that this series is asymptotic but that 
it is not Borel summable. Nonetheless,  we will argue in section 5 that
 the leading terms can be re-summed to an integral representation which is analytic in the
region of interest.  The absence of Borel summability means that such a
 function
is not unique. However it can only differ from the true result by terms which 
have essential singularities at $y=1$ (and vanish there)
 and we do not expect these to affect the 
spectral dimension; so long as there is any power law scaling as $y\to 1$
any vanishing essential singularities are irrelevant.

 Now let
\beq{n3} h_{\RB}^{(n)}=\diff^n h_\RB(y)\yy.\eeq
Then 
\beq{n4}-\diff^{n+1}\frac{1}{h_\RB(y)}\yy=\sum_{r=1}^{n+1}\frac{(-1)^{r+1}r!}
{ (h_{\RB}^{(0)})^{r+1}}\sum_{\C F(n+1,r)}(n+1;a_1,\ldots a_{n+1})'
\prod_{i=1}^{n+1}(h_{\RB}^{(i)})^{a_i}\eeq
where $(m;a_1\ldots a_m)'$ is the multinomial coefficient \cite{Abramowitz}
\beq{n4a}(m;a_1\ldots a_m)'=\frac{m!}{(1!)^{a_1}\ldots(m!)^{a_m}
a_1!\ldots a_m!}\eeq
  and the summation region
$\C F(n+1,r)$ consists of all non-negative integers $a_{1},\ldots, a_{n+1}$ such that 
\beq{n5} \sum_{k=1}^{n+1}ka_k=n+1,\quad \sum_{k=1}^{n+1}a_k=r.\eeq
Inserting \rf{n4} into \rf{n1} and reordering the sums gives
\bea\Qt_n(z)&=&\sum_{r=1}^{n+1}{(-1)^{r+1}r!}\sum_{\C F(n+1,r)}(n+1;a_1,\ldots a_{n+1})'
\left(\sum_{\RB\in\C B} z^{N_\RB}\frac{\prod_{i=1}^{n+1}(h_{\RB}^{(i)})^{a_i} }{ (h_{\RB}^{(0)})^{r+1}}\right).\label{n6}\eea
Let us define
\beq{n7a} H^{(n_1,n_2,\ldots,n_p)}(z)=\sum_{\RB\in\C B}z^{N_B}
\prod_{i=1}^ph_{\RB}^{(n_i)},\qquad p,n_i>0.\eeq
We will prove in section 5 that
\beq{n7} H^{(n_1,n_2,\ldots,n_p)}(z)\simeq A^{(n_1,n_2,\ldots,n_p)}
(1-4z)^{\half-p-\threehalves\sum_{i=1}^p n_i}
\eeq
where the symbol $\simeq$ means the most singular part as
$z\uparrow\quarter$ and  where $ A^{(n_1,n_2,\ldots,n_p)}$ is a constant.
Using \rf{3.19} we note that 
\bea H^{(a_1\otimes 1,a_2
\otimes 2,\ldots,a_{n+1}\otimes n+1)}(z)&=& \left(2z\frac{\partial~}{\partial z}-1\right)^{r+1}
\left(\sum_{\RB\in\C B} z^{N_\RB}\frac{\prod_{i=1}^{n+1}(h_{\RB}^{(i)})^{a_i} }{ (h_{\RB}^{(0)})^{r+1}}\right)\label{n8}
\eea
where by $q\otimes m$ we mean that the number $m$ appears $q$ times 
in the list.
Using \rf{n7} and 
integrating $r+1$ times we find the leading singular behaviour of $\Qt_n(z)$
for $n\ne 0$ to be
\bea \Qt_n(z)&\simeq&(1-4z)^{-\threehalves n}\sum_{r=1}^{n+1}
\frac{(-1)^{r+1}r!}{2^{r+1}}
\frac{\Gamma(\threehalves n)}{\Gamma( \threehalves n+r+1)}\nn\\
&&\times  \sum_{\C F(n+1,r)}(n+1;a_1,\ldots a_{n+1})A^{(a_1\otimes 1,a_2
\otimes 2,\ldots,a_{n+1}\otimes n+1)}.\label{n9}
\eea
  When $n=0$  we obtain
\bea \Qt_0(z)&=&-\eighth\log(1-4z).
\label{n10}
\eea
Substituting \rf{n9} and \rf{n10} in \rf{n2} we see that the most singular part of $\Qt(z,y)$ as $z\uparrow\quarter$ is the sum of
a logarithmic piece and a function of $(1-y)(1-4z)^{-\threehalves}$. For the rest of the
analysis it is more convenient to work with 
$\frac{\partial\Qt}{\partial z}$ which takes the form
\beq{n11} \frac{\partial\Qt}{\partial z}=
\frac{1}{1-4z}\Pt\left(\frac{1-y}{(1-4z)^\threehalves}\right)
\eeq
for some function $\Pt$. From now on in this paper we will be working with
this most singular (as $z\uparrow\quarter$) part of $\Qt$ unless otherwise
stated. Note that from \rf{n11} we deduce that the gap exponent $\Delta=\threehalves$.

\subsection{Extracting $d_S$ from the GCE}

The behaviour of $\Qt(z,1)=\Qt_0(z)$ given by \rf{n10} can be used to place an
upper bound
on $d_S$ which we will need later. From \rf{3.4}, \rf{3.16}, \rf{3.21} we have
\beq{ff1}\Qt(z,1)=\sum_N N^{-\threehalves}(4z)^N\Pb'_N(1).\eeq
Together with \rf{n10} this implies that $\Pb'_N(1)\sim N^\half$ at large
$N$. But from \rf{ii4a} this can only happen if $d_S<2$.

As we discussed in section 2,  the GCE generating function is related to the
  CE generating function by
\bea \frac{\partial\Qt}{\partial z}&=&\frac{1}{u}\Pt\left(\frac{1-y}{u^\Delta}\right)
\nn\\
&=&\int_0^\infty dx\,x^{\gstr-1}\,\Phi(1-y,x)\exp(-xu)\label{f1}
\eea
where, for convenience,  we have introduced  $u=1-4z$ and a 
continuous variable $x$ equivalent at integer values
to $N$, the size of the system. Since we are looking for the asymptotic behaviour
at large $N$ we can approximate the discrete sum over $N$ by a Laplace Transform in
$x$; the difference will be sub-leading.  The extra factor of $x$ appears in \rf{f1} compared to \rf{ii7}
because we are looking at $\frac{\partial\Qt}{\partial z}$ rather than $\Qt$;
 this makes life easier because it removes the possibility of a trivial divergence
at small $x$. From \rf{f1} it follows that 
\beq{f2}\Phi(1-y,x)=x^{1-\gstr}\phi((1-y)x^\Delta)\eeq
for a suitable function $\phi$.
To show this introduce new variables $u'=u(1-y)^{-1/\Delta}$  and $x'=x(1-y)^{1/\Delta}$ into \rf{f1}
to get
\bea\frac{1}{u'}\Pt\left(\frac{1}{u'^\Delta}\right)
=(1-y)^{-(\gstr-1)/\Delta}\int_0^\infty dx'\,x'^{\gstr-1}\Phi(1-y,
\frac{x'}{(1-y)^{1/\Delta}})\exp(-x'u')\label{f3}
\eea
but by setting $y=0$ and  $u=u'$ in \rf{f1} we also have
\beq{3a}\frac{1}{u'}\Pt\left(\frac{1}{u'^\Delta}\right)
=\int_0^\infty dx\,x^{\gstr-1}\,\Phi(1,x)\exp(-xu').
\eeq
Similar identities are valid for the derivatives of $\Pt$ w.r.t. $z$ so
\beq{f4} \Phi(1,x)=(1-y)^{-(\gstr-1)/\Delta}\Phi(1-y,
\frac{x}{(1-y)^{1/\Delta}})
\eeq
and the result \rf{f2} follows.
Now it is easy to see that $\Phi(1-y,x)$ must be a bounded quantity 
in $x>0$ for fixed $y<1$; we have
\bea \Phi(1-y,N)=\frac{1}{T_N}\sum_{i=1}^{T_N}\sum_{t=0}^\infty y^t\left(P_i(t)-\frac{1}
{2N-1}\right)\label{f5}
\eea
where the subtraction is to remove the pole term.
Now $P_i(t)<1$ and $P_i(0)=1$ so we get
\bea 1-\frac{1}{2N-1}< \Phi(1-y,N) <\frac{1}{1-y}\left(1-\frac{1}{2N-1}\right).\label{f6}
\eea
It follows from \rf{f6} that
$ \lim_{x\to\infty}\Phi(1-y,x)$ must be finite for $y<1$ and this is compatible
with \rf{f2} only if $\phi(s)$ behaves like,
\beq{f7} \phi(s)\stackrel{ s\to\infty}{\sim}  s^{-(1-\gstr)/\Delta}\eeq
and so we conclude that
\beq{f8}\lim_{x\to\infty}\Phi(1-y,x)\sim (1-y)^{-(1-\gstr)/\Delta}.
\eeq
Substituting in the values of $\gstr=\half$ and $\Delta=\threehalves$ for the branched polymer ensemble $\C B$ we
obtain
\beq{f9}\lim_{x\to\infty}\Phi(1-y,x)\sim (1-y)^{-\third}.\eeq
This is the generating function $\lim_{N\to\infty}\Pb'_N(y)$  so we find
that  $\lim_{N\to\infty}P_N(t)\sim t^{-\twothirds}$  
and hence that  $d_S=\fourthirds$ (see \rf{ii3} and \rf{ii4}).

\new{}
\subsection{Sub-leading terms}
There are two sources of subleading behaviour. The first is the corrections
to the asymptotic form \rf{f7} which must be like
\beq{f10}\phi(s)\stackrel{ s\to\infty}{\sim}  s^{-(1-\gstr)/\Delta}
\left(1+\frac{const}{s^\epsilon}\right),\quad\epsilon>0\eeq
in order for \rf{f6} to be satisfied.  The correction term is more singular 
as $y\uparrow 1$, and so has smaller $d_S$, but is suppressed by
$N^\epsilon$ in the thermodynamic limit.  This is expected because we know
that there are members of $\C B$ (for example the linear polymers
which are basically  infinite chains with finite sized outgrowths) 
which have $d_S=1$ but are thermodynamically insignificant.

The second source of subleading behaviour is the corrections to \rf{n7}
which, as we will show in section 5, take the form
$u^{\delta+\half-p-\threehalves\sum_{i=1}^p n_i}$ with $\delta=\half,1,\threehalves\ldots$. The consequence of this is that \rf{n11} is modifed to
\beq{f11} \frac{\partial\Qt}{\partial z}=
\frac{1}{1-4z}\Pt\left(\frac{1-y}{(1-4z)^\threehalves}\right)
+\sum_i \frac{1}{(1-4z)^{1-\delta_i}}\Pt_i\left(\frac{1-y}{(1-4z)^\threehalves}\right)
\eeq
where the functions $\Phi_i(1-y,N)$ have the same upper bound as
$\Phi(1-y,N)$, see \rf{f6}.
Repeating the manipulations of section 4.2, and assuming that the $i$'th term
in \rf{f11} makes a contribution which survives as $N\to\infty$ we find
by the arguments of the previous sub-section that
it has a spectral dimension
\beq{f12} d_S^i=2+\twothirds(2\delta-1)\ge 2.\eeq
This shows that it is not possible that the leading term, i.e. the $\Pt$ term,
 makes no
contribution as $N\to\infty$ and that the lower  bound \rf{f6} is fulfilled
by one of the subleading terms; this would lead to $d_S\ge 2$ for $\C B$
in violation of the bound $d_S< 2$ obtained from the behaviour at $y=1$.
\new

\section{ Proof of the Principal Result}

In this section we will prove  the result \rf{n7} and derive constraints
on the coefficients $A^{(n_1,\ldots n_p)}$. It is convenient for what follows
to adopt a more concise notation. Let $S$ denote a list of non-negative integers
$n_1,\ldots n_{p_S}$ and let
\beq{p0} n_{S}=\sum_{n_i\in S}n_i.\eeq
We adopt the convention that the empty list corresponds to $p_S=0$.
In this notation we will prove that leading singular behaviour
as $z\uparrow\quarter$ is given by
\bea H^{(S)}(z)&=&\sum_{\RB\in\C B}z^{N_B}
\prod_{i=1}^{p_S}h_{\RB}^{(n_i)},\qquad p_S,n_i>0,\nn\\
&\simeq& A^{(S)}u^{\half-p_S-\threehalves n_S},\qquad u=1-4z,
\label{p1}
\eea
where $ A^{(S)}$ is a constant. We will refer to the 
power of $u$ in \rf{p1} as the \emph{actual degree} (of singularity)
of $ H^{(S)}(z)$.  If $p_S=0$ we have the special 
case
\beq{p2}
H^{()}(z)={\C T}(z)\simeq\half.
\eeq
We note that, having
established  \rf{p1} for a given set of integers $S$, it is a
 corollary of \rf{p1} that 
\bea H^{(q\otimes 0,S)}(z)&=&\sum_{\RB\in\C B}z^{N_B}
\left(h^{(0)}_{\RB}\right)^q\prod_{i=1}^{p_S}h_{\RB}^{(n_i)},\qquad n_i>0,\nn\\
&\simeq& A^{(S)} 2^q\frac{\Gamma(\threehalves n_S  +p_S+q-\half)}
{\Gamma(\threehalves n_S+p_S-\half)}
u^{\half-p_S-q-\threehalves n_S}.
\label{p3}
\eea
This follows from \rf{3.19} since
\bea H^{(q\otimes 0,S)}(z)&=&\sum_{\RB\in\C B}z^{N_B}
\left(2N_{\RB}-1\right)^q\prod_{i=1}^ph_{\RB}^{(n_i)}\nn\\
&=&\left(2z\frac{\partial~}{\partial z}-1\right)^q H^{(S)}(z)
\nn\\
&\simeq& \left(\half\,\frac{\partial~}{\partial z}\right)^q A^{(n_1,n_2,\ldots,n_p)}
u^{\half-p_S-\threehalves n_S}
\label{p4}
\eea
and the result follows.

We start by computing the  derivatives for polymer  $\RA=\RB\cup\RC$ in 
terms of those for its constituents $\RB$ and  $\RC$; differentiating
 \rf{3.14} $n$ times and setting $y=1$ we obtain
\bea
 h_\RA^{(n)}&=&h_\RB^{(n)}+h_\RC^{(n)}\nn\\
&&+\sum_{r=1}^{n}\frac{n!}{n-r!}
\diff^{n-r}\Bigg\{(h_\RB(y))^r(1+h_\RB(y)) +\BtoC \nn\\
&&\qquad\qquad\qquad\qquad+\sum_{k=1}^{r-1}{r\choose k}(h_\RC(y))^k(h_\RB(y))^{r-k}+\nn\\
&&\qquad\qquad\qquad\qquad+\sum_{k=1}^{r}{r+1\choose k}(h_\RC(y))^k(h_\RB(y))^{r+1-k}\Bigg\}\yy.
\label{p5}
\eea
Consider first the case $p=1$, $n_1=1$. We have
\beq{p6}
 h_\RA^{(1)}=h_\RB^{(1)}+h_\RC^{(1)}+h_\RB^{(0)}(1+h_\RB^{(0)})
+h_\RC^{(0)}(1+h_\RC^{(0)})+2h_\RB^{(0)}h_\RC^{(0)}.
\eeq
Summing over all $\RA=\RB\cup\RC$ (and remembering that $h_{\RB_0}^{(1)}=0$)
we find that 
\bea
 H^{(1)}(z)&=&\sum_{\RA\in\C B}z^{N_A}h_\RA^{(1)}\nn\\
&=&\left\{\sum_{\RB\in\C B}z^{N_B}\left(h_\RB^{(1)}+h_\RB^{(0)}(1+h_\RB^{(0)})\right)
\sum_{\RC\in\C B}z^{N_C}+\BtoC\right\}\nn\\
&&\qquad+2\sum_{\RB\in\C B}z^{N_B}h_\RB^{(0)}\sum_{\RC\in\C B}z^{N_C}h_\RC^{(0)}\nn\\
&=&2{\C T}(z)\left( H^{(1)}(z)+H^{(0,0)}(z)+H^{(0)}(z)\right)+
2H^{(0)}(z)H^{(0)}(z).
\label{p7}
\eea
This can be rearranged to give, using \rf{p1} and \rf{p3}, 
\beq{p8}
 H^{(1)}(z)(1-2{\C T}(z))=\half(1-4z)^{-\threehalves}+O((1-4z)^{-1}).
\eeq
Note that the most singular term on the r.h.s. of \rf{p8} comes from
the $H^{(0,0)}(z)$ term in \rf{p7}. Finally, substituting for ${\C T}(z)$ from
\rf{3.5}, we find
\beq{p9}
 H^{(1)}(z)\simeq\half(1-4z)^{-2}
\eeq
in accordance with \rf{p1}.

We can now proceed to the general case by induction. The strategy is always the same
as the one we have just used. We take the definition
\beq{p10} H^{(S)}(z)=\sum_{\RA\in\C B}z^{N_A}
\prod_{i=1}^{p_S}h_{\RA}^{(n_i)}\eeq
and replace $h_{\RA}^{(n_i)}$ by its expression \rf{p5} in terms of 
its constituent polymers $\RB$ and $\RC$. By doing this for 
sets of integers $S$ chosen in a particular order 
 we can ensure that the resulting expression on the
r.h.s. of \rf{p10} contains only $ H^{(\tilde S)}(z)$
which are already known. To see this we can write \rf{p5}
as
\beq{p10a}  h_\RA^{(n)}=h_\RB^{(n)}+h_\RC^{(n)}+T(n-1)\eeq
where by $T(n-1)$ we denote all contributions which 
contain  derivatives of  no higher order than $n-1$. Let 
$S=(n_1,n_2,\ldots,n_{P_S})$ where each $n_i\le n-1$. Then  consider
\bea
 H^{(q\otimes n,S)}(z)&=&
\sum_{\RB\in\C B}z^{N_B}\sum_{\RC\in\C B}z^{N_C}
\left(h_\RB^{(n)}+h_\RC^{(n)}+T(n-1)\right)^q\nn\\
&&\quad\quad\quad\quad\times\prod_{i=1}^{p_S}\left(h_\RB^{(n_i)}+h_\RC^{(n_i)}+T(n_i-1)\right).
\label{p10c}
\eea
 After multiplying out and doing the
sums over B and C the generic term on the r.h.s. of \rf{p10c} 
is of the form
\beq{p10d} H^{(S_1)}(z) H^{(S_2)}(z)\eeq
where $S_{1}$ and $S_{2}$ are lists of integers with the property that no
 member is higher than $n$.  We will denote by  $S_-$ a generic
list of numbers derived
from $S$ by removing one element and by $S_+$ a generic 
list of numbers derived
from $S$  by removing one element
$n_k$ and replacing it by $\{m_1,...\}$ where the $m_i$ add up to $n_k$.
We can categorize the possible pairs of lists $S_{1}$ and $S_{2}$ 
that appear in \rf{p10d}
 by keeping track of the number
of factors of $T$ which appear on multiplying out.
\begin{enumerate}
\item No $T$ factors. Apart from the two terms which reproduce the l.h.s.
multiplied by $\C T(z)$ these give $S_{1}$ and $S_{2}$ each with either
a) fewer than $q$ elements with the value $n$ and all other elements
 less than $n$ (since they are drawn from $S$)
or
b) $q$ elements with the value $n$ but fewer than $p_S$ remaining elements
  less than $n$ drawn from  $S$.

\item One or more $T(n-1)$ factors and any number (up to $p_S$) of $T(n_i-1)$
factors. In this case each of $S_{1}$ and $S_{2}$  contain $n$ fewer than $q$ times with all other members
being less than $n$.

\item  One or more $T(n_i-1)$ factors and no $T(n-1)$
factors. In this case $S_{1}$ and $S_{2}$
 contain $n$ up to $q$ times, at most $p_S-1$ members 
drawn from $S$ and the rest are drawn from the integers $\le n-2$.

\end{enumerate}

\noindent From 1a and 2 we see that $H^{(q-1\otimes n,S)}(z)$ must be 
calculated before $H^{(q\otimes n,S)}(z)$ and from 
1b that $H^{(q\otimes n,S_-)}(z)$ must be 
calculated before $H^{(q\otimes n,S)}(z)$.  Finally, 
from 3 we see that $H^{(q\otimes n,S_+)}(z)$
must be computed before $H^{(q\otimes n,S)}(z)$. These criteria
are automatically satisfied by lists of numbers ordered in the following way
(below $S_a>S_b$ means that $H^{(S_b)}$ is calculated before $H^{(S_a)}$).
Let  the largest number appearing in the list $S_{a}$ be
$n_{a}$ and denote the number of  appearances by $q_{a}$. The ordering
is defined as follows:
\begin{enumerate}
\item If $n_a>n_b$ then $S_a>S_b$.
\item If $n_a=n_b$, and $q_a>q_b$ then $S_a>S_b$.
\item If $q_a=q_b$ then replace $n_{a}$ and $n_{b}$ by the next
 largest numbers
appearing in $S_{a}$ and $S_{b}$ respectively
 and apply 1.,2., and 3. recursively until the
issue is decided (if there are no more numbers in a list then the
corresponding $n$ is set to zero).
\end{enumerate}
\noindent Of course to calculate as far as a given $S$ it is not 
necessary to know the result for all preceding lists in the ordering 
defined above
because \rf{p10c} only requires knowledge of preceding lists with at most $n_S+1$
elements.

Now we need to keep track of the degree of singularity of the terms
\rf{p10d} appearing on the r.h.s. of \rf{p10c}.
Define the \emph{ naive degree } of $h_{\RB}^{(n_i)}$ as
\beq{p11} d_0(h_{\RB}^{(n_i)})=-1-\threehalves n_i\eeq
and the naive degree of a product of such terms
\beq{p12} d_0(\prod_{i=1}^p h_{\RB}^{(n_i)})=\sum_{i=1}^p 
d_0(h_{\RB}^{(n_i)}).
\eeq
Then, if \rf{p1} is already  established  for the lists  $S_{1}$ and 
$S_{2}$,
the actual degree is given by
\beq{p13} d=1+d_0,\quad\mathrm{(Case~1)}\eeq
\emph{unless} it happens that the product consists entirely of
$h_{\R B}^{(n_i)}$ and no $h_{\R C}^{(n_i)}$ (or vice versa) in which case
\beq{p14} d=\half+d_0,\quad\mathrm{(Case~2)}\eeq
(this is because, anomalously, $H^{()}\simeq\half$ and not $u^\half$).
 Now compare two terms of naive degree $d_0'$ and $d_0''$ respectively which 
satisfy $d_0'<d_0''-\half$; 
then the  corresponding actual degrees $d'$ and $d''$ satisfy $d'<d''$ 
 because
\beq{p15} \max(d')=d_0'+1<d_0''+\half=\min(d'').\eeq
This result can be used to discard terms in \rf{p10}
 that can only make 
sub-leading contributions to $ H^{(S)}(z)$; only terms whose
naive degree is at most $\half$ greater than the lowest naive degree  appearing
in an expression can possibly contribute to the leading divergence at 
$z\uparrow\quarter$. We will call this result the ``naive degree criterion''.

To keep account of the naive degree in subsequent formulae,
we will adopt the notation that quantities in square brackets $[~]$
immediately after a term denote its value of $d_0$. Anotating \rf{p5}
in this way
\bea
 h_\RA^{(n)}&=&h_\RB^{(n)}[{\scriptstyle -1-\threehalves n}]+\BtoC\nn\\
&&+\sum_{r=1}^{n}\frac{n!}{n-r!}
\diff^{n-r}\Bigg\{(h_\RB(y))^r[{\scriptstyle \half r-\threehalves n}]
+(h_\RB(y))^{r+1}[{\scriptstyle \half r-1-\threehalves n}] +\BtoC\nn\\
&&\qquad\qquad\qquad\qquad+\sum_{k=1}^{r-1}{r\choose k}(h_\RC(y))^k(h_\RB(y))^{r-k}[{\scriptstyle \half r+1-\threehalves n}]\nn\\
&&\qquad\qquad\qquad\qquad+\sum_{k=1}^{r}{r+1\choose k}(h_\RC(y))^k(h_\RB(y))^{r+1-k}[{\scriptstyle \half r-\threehalves n}]\Bigg\}\yy.
\label{p16}
\eea
Using the naive degree criterion  we see immediately that there is only
one term in the braces which can ever contribute to the leading divergence
which can therefore be computed from the expression
\bea 
 H^{(S)}(z)&\simeq&
\sum_{\RB\in\C B}z^{N_B}\sum_{\RC\in\C B}z^{N_C}\prod_{i=1}^{p_S}
\Bigg(h_\RB^{(n_i)}[{\scriptstyle -1-\threehalves n}]\nn\\
&&+n_i\diff^{n_i-1}
h_\RB(y)^2\yy[{\scriptstyle -\half-\threehalves n}]+\BtoC\Bigg).
\label{p17}\eea
When we multiply out the r.h.s. of \rf{p17} 
the naive degree criterion again tells us that only
contributions containing at most one factor of the 
term with $d_0=-\half-\threehalves n$ need be retained leaving
\bea  H^{(S)}(z)(1-2\C T(z))&\simeq&
\sumprime_{\scriptstyle S_1\cup S_2=S} H^{(S_1)}(z)
H^{(S_2)}(z)\nn\\
&&+2\C T(z)\sum_{k=1}^{p_S} \sum_{m=0}^{n_k-1}n_k{n_k-1\choose m}
H^{(n_k-1-m,m, S\backslash n_k)}(z)
\label{p18}\eea
where the primed sum indicates that the empty list is not included,
 and $S\backslash n_k$ is the list $S$ with the number $n_k$ removed.
Both terms on the r.h.s. of \rf{p18} have actual degree 
$1-p_S-\threehalves n_S$ and so the result \rf{p1} follows 
for $ H^{(S)}(z)$.

Having established the leading behaviour of $H^{(S)}$ it is straightforward
to categorize the possible sub-leading behaviour. From \rf{p18} we note that 
one contribution to the first sub-leading term  of $H^{(S)}$ arises when on the r.h.s. we take the first sub-leading term  of $H^{(S_1)}$ and 
the leading term  of $H^{(S_2)}$ (or vice versa). Assuming that 
the first sub-leading term  of $H^{(S)}$ takes the form
$u^{a-bp_S-cn_S}$ and equating powers we get
\beq{p18a}u^\half u^{a-bp_S-cn_S}\simeq
 u^{\half-p_{S_2}-\threehalves n_{S_2}}u^{a-bp_{S_1}-cn_{S_1}}\eeq
which immediately tells us that $b=1$, $c=\threehalves$. From \rf{p16} and 
the  result \rf{p8} for $H^{(1)}$    we also see that the first 
 sub-leading term is just $u^\half$ less singular than the leading one.
Repeating this argument for lower divergences we conclude that a general
sub-leading contribution to  $H^{(S)}$ must take the form
\beq{p18b}  u^{\delta+\half-p_{S}-\threehalves n_{S}},\quad
\delta=\scs\half,1,\scs\threehalves,\ldots\,.\eeq

\new
We can also use \rf{p18} to   obtain  an expression
relating the coefficients $A^{(S)}$ 
\beq{p19}  A^{(S)}=\sumprime_{\scriptstyle S_1\cup S_2=S}A^{(S_1)}
A^{(S_2)}
+\sum_{k=1}^{p_S} \sum_{m=0}^{n_k-1}n_k{n_k-1\choose m}
A^{(n_k-1-m,m, S\backslash n_k)}.
\eeq
Although these recursion relations cannot be solved exactly we can use them
to obtain estimates of  the coefficients $ A^{(S)}$. First we
show that all $ A^{(S)}$ are positive.  Assume
that the  $ A^{(S)}$ are computed in the sequence described above.
From \rf{p3} we see that if
$ A^{(\tilde S)}$ is positive for some $\tilde S$ then so is 
$ A^{(q\otimes0,\tilde S)}$ and it 
follows that if $ A^{(\tilde S)}>0$ then also $ A^{(\tilde S_+)}>0$.  Since
$A^{(1)}>0$ (from \rf{p9}) it follows that all  $ A^{(S)}$ are positive.
It is convenient to  define a new coefficient $B^{(S)}$, which is also always positive, by
\beq{p20} A^{(S)}= B^{(S)}\prod_{i=1}^{p_S}n_i!\eeq
and substitute this in \rf{p19}; all the factorials cancel leaving 
\beq{p21}  B^{(S)}=\sumprime_{\scriptstyle S_1\cup S_2=S}B^{(S_1)}
B^{(S_2)}
+\sum_{k=1}^{p_S} \sum_{m=0}^{n_k-1}
B^{(n_k-1-m,m, S\backslash n_k)}.
\eeq

We can obtain a lower bound on $B^{(S)}$ by introducing $\tilde B^{(S)}$ which 
satisfies $\tilde B^{()}=B^{()}$, $\tilde B^{(q\otimes 0)}=B^{(q\otimes 0)}$,
and
\bea  \tilde B^{(S)}=\sum_{k=1}^{p_S} \sum_{m=0}^{n_k-1}
\tilde B^{(n_k-1-m,m, S\backslash n_k)}\label{p21a}\eea
which is \rf{p21} with the quadratic term dropped.
Now \rf{p21a} is solved by $\tilde B^{(S)}=b(n_S,p_S)$ where
\bea b(n,p)&=&n\, b(n-1,p+1).
\label{p22}\eea
Iterating this equation we find that
\beq{p23} b(n,p)=n!\,b(0,p+n).\eeq
But $b(0,p+n)=B^{((p+n)\otimes 0)}$ which we can compute exactly from \rf{p3},
 so we obtain
\beq{p24}  b(n,p)=2^{n+p-2}n!\,\frac{\Gamma(n+p-\half)}{\Gamma(\half)}.\eeq
Now note from \rf{p3} and \rf{p24}  that
\beq{p24a}\frac{B^{(S,q\otimes 0)}}{B^{(S)}}>\frac{b(n_S,p_S+q)}{b(n_S,p_S)}.\eeq
This ensures that those terms on the r.h.s. of \rf{p21} and
\rf{p21a} which arise when
$m=0$ or $n_k=1$ are larger for the true coefficients $B^{(S)}$ than they are
for the $\tilde B^{(S)}$. It now follows that,
since \rf{p21a} is just \rf{p21} without the quadratic term,  
\beq{p24b} B^{(S)} > b(n_S,p_S).\eeq
Finally note that, as a consistency check, for $p=1$, $n_1=1$,
 this lower bound evaluates to give
 $A^{(1)}=\half$ which is the exact result \rf{p9}.

To find an upper bound on $B^{(S)}$ we first  need the subsidiary
 result that
\bea Q&=&\sumprime_{\scriptstyle S_1\cup S_2=S}
\Gamma(\alpha n_{S_1}+p_{S_1}-\ts\half)
\Gamma(\alpha n_{S_2}+p_{S_2}-\ts\half)\nn\\
&\le&4\Gamma(\ts\half)\Gamma(\alpha n_{S}+p_{S}-\ts\half).\label{p26}\eea
This can be proved by replacing the Gamma functions with their integral
representation and using the inequality $t^{\beta}+s^{\beta}
\le (t+s)^{\beta}$ for $t,s \ge 0$ and $\beta\ge 1$.  Now define
\beq{p27}B_+^{(S)}=B_0\Gamma(\alpha\,n_S+p_S-\ts\half)
r^{\alpha n_S+p_S}\eeq
where $B_0$, $\alpha$, and $r$ are constants. By \rf{p26} we have
\beq{p28}\sumprime_{\scriptstyle S_1\cup S_2=S}B_+^{(S_1)}B_+^{(S_2)}
\le 4\Gamma(\ts\half) B_0 B_+^{(S)}\eeq
and
\bea
\sum_{k=1}^{p_S} \sum_{m=0}^{n_k-1}
B_+^{(n_k-1-m,m, S\backslash n_k)}&=&\left(\sum_{i=1}^{p_S}n_i\right) B_0
\Gamma(\alpha n_S+p_S-\ts\half+(1-\alpha))
r^{1-\alpha} r^{\alpha n_S+p_S}\nn\\
&=&\frac{B_0r^{1-\alpha}}{\alpha}\left(\alpha n_S+p_S-\ts\half+(1-\alpha)
-(p_S-\ts\half+1-\alpha)\right)\nn\\
&&\times \Gamma(\alpha n_S+p_S-\ts\half+(1-\alpha))
 r^{\alpha n_S+p_S}\nn\\
&<&
\frac{B_0r^{1-\alpha}}{\alpha}\Gamma(\alpha n_S+p_S-\ts\half+(2-\alpha))
r^{\alpha n_S+p_S}.\label{p29}
\eea
If we choose $\alpha <2$ then this is generically bigger than
$B_+^{(S)}$ whereas if we choose $\alpha >2$ it is generically smaller.
Combining \rf{p26} and \rf{p29} we find that for $\alpha =2$
\beq{p30}\sumprime_{\scriptstyle S_1\cup S_2=S}B_+^{(S_1)}
B_+^{(S_2)}
+\sum_{k=1}^{p_S} \sum_{m=0}^{n_k-1}
B_+^{(n_k-1-m,m, S\backslash n_k)}<\left(4\Gamma(\ts\half) B_0+\frac{1}{2r}\right)
B_+^{(S)}.\eeq
Now note from \rf{p3} and \rf{p27} that
\beq{p30a}\frac{B^{(S,q\otimes 0)}}{B^{(S)}}<\frac{B_+^{(S,q\otimes 0)}}
{B_+^{(S)}}.\eeq
This ensures that those terms on the r.h.s. of \rf{p21} and which arise when
$m=0$ or $n_k=1$ are smaller for the true coefficients $B^{(S)}$ than they are
for the $ B_+^{(S)}$.
The numbers $B_+^{(S)}$ provide consistent upper bounds on $B^{(S)}$ if we can choose
$B_0$ and $r$ such that the r.h.s. of \rf{p30} is less than $B_+^{(S)}$
and such that $B^{(1)}<B_+^{(1)}$.  It is easy to see that this can be
accomplished by choosing $r=2$ and
\beq{p31}
B_0=\left(\third+\epsilon\right)\frac{1}{4\Gamma(\half)}\eeq
where $\epsilon$ is small and positive so we conclude that
\beq{p32}B^{(S)}<B_0
\Gamma(2 n_S+p_S-\textstyle{\half})
2^{2n_S+p_S}.\eeq

Both the upper \rf{p32} and the lower \rf{p24} bounds on $B^{(S)}$ behave as
$\Gamma(2n_S+p_S-\half)$ with an exponential prefactor.  With coefficients of this
form the Taylor series  for $\Qt (z,y)$ \rf{n2} is re-summable which we will
demonstrate for the upper bound $B_+^{(S)}$.  From \rf{n2} and \rf{n10} we have
\beq{p33}\Qt(z,y)=-\eighth\log(1-4z)+G(z,y).\eeq
Replacing the coefficients $B^{(S)}$ by their upper bounds $B_+^{(S)}$
and using \rf{p20} and \rf{p27} $G(z,y)$ becomes
\bea G_+(z,y)&=&\half\sum_{n=1}^\infty \left(\frac{y-1}{(1-4z)^\threehalves}
\right)^n
\frac{1}{n!}\sum_{r=1}^{n+1}(-1)^{r+1}\frac{r!\Gamma(\threehalves n)}
{\Gamma(\threehalves n+r+1)} 2^{2(n+1)}\nn\\
&&\times \sum_{\C F(n+1,r)}\frac{(n+1)!\Gamma(2( n+1)+r-\half)}{a_1!\ldots a_{n+1}!}\label{p34}.\eea
Using the integral representations of the Beta and Gamma functions and interchanging the order of the summations gives
\bea G_+(z,y)&=& 2\frac{\partial~}{\partial w}\int_0^1\frac{dt}{t}
\int_0^\infty\frac{e^{-s}ds}{s}\Bigg\{(1-t)s^\fivehalves w\sum_{n=1}^\infty
(wt^\threehalves s^2)^n\nn\\
&& +s^{-\half}\sum_{r=2}^\infty (-1)^{r+1}\sum_{m\ge r}
(wt^\threehalves s^2)^m\sum_{\C F(n+1,r)}\frac{1}{a_1!\ldots a_{n+1}!}\Bigg\}
\label{p35}\eea
where
\beq{p35a}w=\frac{y-1}{(1-4z)^\threehalves}.\eeq
The remaining sums can be done leaving the integral representation
\bea G_+(z,y)&=& 2\frac{\partial~}{\partial w}\int_0^1\frac{dt}{t}
\int_0^\infty\frac{e^{-s}ds}{s}\nn\\
&&\Bigg\{(1-t)s^\fivehalves w\frac{wt^\threehalves s^2}{1-wt^\threehalves s^2}-s^{-\half}t^{-\threehalves}\left(e^\zeta-1-\zeta\right)\Bigg\}\label{p36}\eea
where
\beq{p37}\zeta=-s(1-t)\frac{wt^\threehalves s^2}{1-wt^\threehalves s^2}.\eeq
We see that, as claimed in section 4.1, $G_+$ is a function of $w$.
 In the region
of interest, $0\le y< 1$, $0\le z<\quarter$, $w$ is always negative and it is
straightforward to check that the integrals are finite, not only for $G_+$ but also
for all derivatives of $G_+$ with respect to $w$.  Thus $G_+$ is analytic in 
this region; it is a function whose asymptotic series about $y=1$ reproduces
the series for $G$ with the coefficients $B^{(S)}$ replaced by $B_+^{(S)}$.
It is not at all trivial that this is the case; indeed if the coefficients
were to have the behaviour $\Gamma(2n_S+\beta p_S-\half)$ with $\beta>1$ then
the resummation would  lead to a divergent integral.
\new
\section{Discussion}
The value $d_S=\fourthirds$ can be compared with the numerical simulations 
of \cite{AmbE} and in particular fig.10 of that paper.  The data for central 
charge $c=5$ is in good agreement with the analytic result for $\C B$. The 
$c=5$ data also yields a value of $\gstr$ close to $\half$ and a
 very good fit to $d_h=2$ which is the analytic 
result for the Hausdorff dimension of $\C B$. Taken altogether this is very strong evidence that at  $c=5$ the quantum gravity ensemble is 
truly branched-polymer like.  On the other hand our result clearly rules out
the conjecture that $d_h=2d_S$ for a unitary quantum gravity ensemble
(this conjecture is also inconsistent with very high precision numerical
results for $c=-2$ which is a non-unitary theory \cite{AmbC}).   For  values
of $c$ closer to one the numerical results for $d_S$ are some distance from
 $\fourthirds$ just as those for $d_h$ and $\gstr$
 are some distance from the generic branched polymer values.  This
is a familiar phenomenon in these models which seem, at least from the numerical point of view on finite size systems, to have complicated cross-over behaviour from $c<1$ properties to generic large $c$ properties.  This can arise from
competing contributions which obscure the asymptotic large volume behaviour and has been discussed at length in \cite{David2}.  Recently simulations have
been performed to determine the spectral dimension of what is believed to be
a branched polymer phase in four-dimensional euclidean quantum gravity; the results are in excellent agreement with the expectation that $d_S=\fourthirds$ 
\cite{private}.

The calculation presented in this paper shows that $d_S=\fourthirds$ for
the simplest example of a generic branched polymer ensemble. By tracing the
origin of the gap exponent $\Delta$ it is straightforward to see that the
same structure will emerge for any other generic ensemble. The crucial point 
is the factor $(1-2\C T(z)) =(1-4z)^\half$ appearing
 on the l.h.s. of \rf{p18};
 the power $\half$
just arises from the structure of the GCE partition function and is simply
$1-\gamma_{str}$ where, generically, $\gamma_{str}=\half$.
 For the non-generic branched polymer ensembles which can
be produced by allowing vertices of arbitrary order and weighting them in
particular ways \cite{bialas} it is possible to generate different values of
$\gamma_{str}$ and we expect that these ensembles will in general
have $d_S\ne\fourthirds$.

Finally we would like to discuss the connection between our calculation and the
spectral dimension on percolation clusters \cite{Orbach}.  These authors 
originally conjectured that the spectral dimension on percolation clusters at 
criticality is $\fourthirds$ independent of the embedding dimension $D$.
High precision numerical calculations have long since shown that this is not
correct for small $D$ (see \cite{Havlin} for a review).
  However for infinite (or sufficiently large) $D$ this result is believed to
be exact and there are a number of scaling arguments and approximate 
calculations for it \cite{stanley}. 
Now consider the set of
(bond) percolation clusters (PC)
 on a Cayley tree with vertices of order three and 
with the constraint  that the 
bond from the root to the first vertex is occupied (fig.5).
 \begin{figure}[h]
{\epsfxsize=\hsgraph \epsfbox{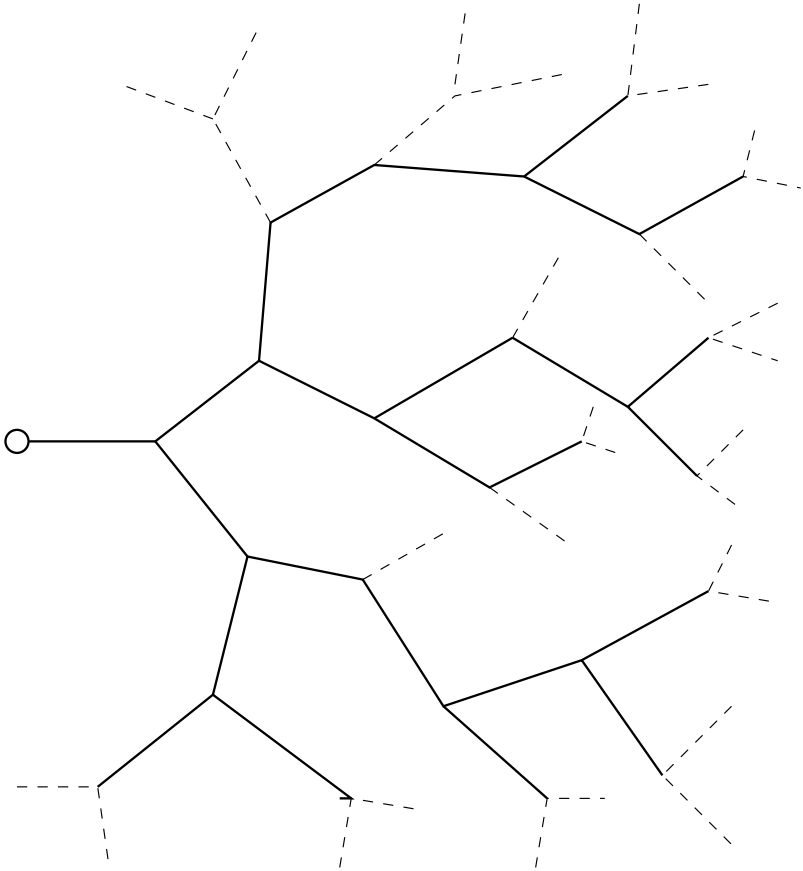}}
\caption{Percolation cluster on a Cayley tree.}
\end{figure}
This set is very similar to the $\C B$ ensemble except that the percolation
clusters contain two-point as well as three-point vertices. It is
 straightforward to map the clusters onto a  branched polymer ensemble which 
contains graphs with two-point and three-point vertices with appropriate
weights. The only subtlety arises in the weighting of the two-point vertex.
Every time a two-point vertex is present in a branched polymer
 there are \emph{two} percolation clusters corresponding to it (fig.6).
 \begin{figure}[h]
{\epsfxsize=\hsgraph \epsfbox{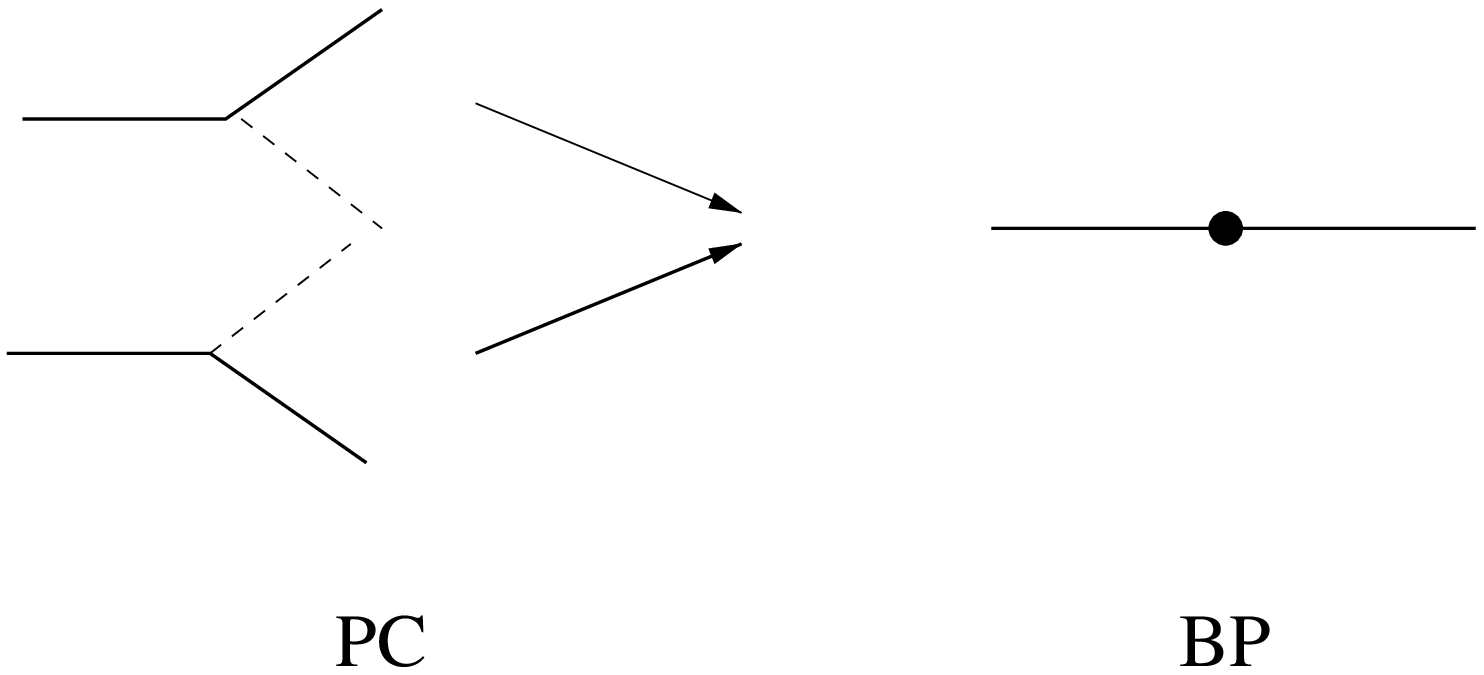}}
\caption{Mapping of percolation cluster two-point vertex to branched polymer.}
\end{figure}
 Thus 
the branched polymer ensemble which generates the percolation clusters with
the correct weights has GCE partition function ${\C Z}$  satisfying (see fig.7)
 \begin{figure}[h]
{\epsfxsize=\hsgraph \epsfbox{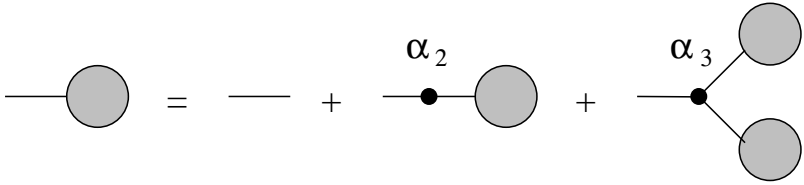}}
\caption{The  branched polymer equation that generates percolation clusters.}
\end{figure}
\beq{pcc} {\C Z}=z(1+\alpha_2{\C Z}+\alpha_3{\C Z}^2)\eeq
where $\alpha_3=1$ but $\alpha_2=2$ and $z$ is the bond occupation probability
in PC language. This modified branched polymer ensemble still has $\gstr=\half$
and hence, according to our discussion above, $d_S=\fourthirds$; in fact it is
the simplest extension of $\C B$.
Because of our interest in quantum gravity we defined $d_S$
 in terms of probabilities averaged over all clusters of size $N$ 
 rather than in terms of a
particular ``typical'' cluster. The calculation picks out the leading 
non-analyticity as $y\to 1$ \emph{after}  taking the thermodynamic limit; we
have \rf{f8}
\beq{d1} \lim_{N\to\infty}\Phi(1-y,N) \simeq \frac{1}{(1-y)^{1-d_S/2}}=\frac{1}{(1-y)^{1/3}}\eeq
This shows that in the thermodynamic limit a finite proportion of the
 $T_N$ clusters of size $N$ have
 $d_S=\fourthirds$.
As we discussed in section 4.3 clusters with $d_S$ \emph{smaller}
than  $\fourthirds$ are present   but the number of such clusters
is suppressed by a power of $N$ relative to $T_N$ and so the probability that
they appear as an infinite percolation cluster is zero.

\vspace{1 truecm}
\noindent T.J. would like to thank Bergfinnur Durhuus for many helpful discussions.
This work was supported in part by PPARC visiting fellowship grant GR/K95086
and by EU grant ERBCHRX CT930343.

\section*{Appendix}
In this appendix we prove that the GCE generating function \rf{3.16} with
the pole term at $y=1$ removed has an asymptotic expansion about $y=1$.

The return probability $P_{\R B}(t)$ on an individual polymer B can be expressed in
terms of a transfer matrix $T_{\R B}=1-D_{\R B}$ where $D_{\R B}$ is the 
Laplacian on B. $D_{\R B}$  has eigenvalues $\{\lambda_{\R B}^i,\;i=1\ldots N_{\R B}\}$; one eigenvalue, $\lambda_{\R B}^{N_{\R B}}$, is zero while the rest satisfy $1\ge \lambda_{\R B}^i  >0$. We have
\beq{one} P_{\R B}(t)=\sum_{i=1}^{{N_{\R B}}}a^i_{\R B}(1-\lambda_{\R B}^i)^t\eeq
where the constants $a^i_{\R B}$ depend on the eigenvectors of $D_{\R B}$
and satisfy $0\le a^i_{\R B}\le 1$. Substituting \rf{one} into \rf{3.16}
but omitting the zero eigenvalue (which generates the pole at $y=1$) we obtain
\beq{two} {\C Q}'(z,y)=\sum_{\R B\in\C B} z^{N_{\R B}}\sum_{t=0}^\infty
y^t\sum_{i=1}^{{N_{\R B}-1}}a^i_{\R B}(1-\lambda_{\R B}^i)^t.\eeq
${\C Q}'$ is absolutely convergent for $0\le y <1$, $0\le z <\quarter$.
Using the binomial expansion we get
\bea\delta{\C Q}'(z,y)&=&{\C Q}'(z,y)-{\C Q}'(z,1)\nn\\&=&
\sum_{\R B\in\C B} z^{N_{\R B}}\sum_{t=0}^\infty\sum_{n=1}^t {t\choose n} (y-1)^n
\sum_{i=1}^{{N_{\R B}-1}}a^i_{\R B}(1-\lambda_{\R B}^i)^t\label{three}.\eea
The term $(y-1)^n$ can be replaced using the integral representation
\beq{four} x^n=\frac{1}{\Gamma(n)}\int_0^\infty v^{n-1} e^{-\frac{v}{x}}\, dv
\eeq
so that, by re-arranging and re-summing over $n$ and $t$, we get
\beq{five}\delta{\C Q}'(z,y)=\int_0^\infty\beta(v)\,e^{-\frac{v}{1-y}}\,dv\eeq
where
\beq{six}\beta(v)=-\sum_{\R B\in\C B} z^{N_{\R B}}
\sum_{i=1}^{{N_{\R B}-1}}a^i_{\R B}\frac{(1-\lambda_{\R B}^i)}{(\lambda_{\R B}^i)^2} \exp\left(-v\frac{1-\lambda_{\R B}^i}{\lambda_{\R B}^i}\right).\eeq
We can now use \rf{five} to develop a formal series in powers of $(1-y)$
by repeatedly integrating by parts which gives
\beq{seven} {\C Q}'(z,y)=
{\C Q}'(z,1)+(1-y)\beta(0)+(1-y)^2\beta'(0)+\ldots+(1-y)^n\beta^{(n-1)}(0)
+R_n\eeq
where
\beq{eight}R_n=(1-y)^n\int_0^\infty\beta^{(n)}(v)\,e^{-\frac{v}{1-y}}\,dv.\eeq
To compute the $n$th derivative of $\beta(v)$ from \rf{six} is trivial
and so is the remaining $v$ integral in \rf{eight} (remember that $y<1$). We
find the remainder term
\beq{nine}R_n=(1-y)^n(-1)^{n+1}\sum_{\R B\in\C B} z^{N_{\R B}}
\sum_{i=1}^{{N_{\R B}-1}}a^i_{\R B}\frac{(1-\lambda_{\R B}^i)^{n+1}(\lambda_{\R B}^i)^{-n-2}}{\frac{1}{1-y}+
\frac{1-\lambda_{\R B}^i}{\lambda_{\R B}^i}} \eeq
and the coefficient
\beq{ten}\beta^{(n)}(0)=(-1)^{n+1}\sum_{\R B\in\C B} z^{N_{\R B}}
\sum_{i=1}^{{N_{\R B}-1}}a^i_{\R B}{(1-\lambda_{\R B}^i)^{n+1}(\lambda_{\R B}^i)^{-n-2}}.\eeq
We see immediately that 
\beq{eleven}R_n=(1-y)^{n+1}(-1)^{n+1}r_n\eeq
where
\beq{twelve}r_n<\beta^{(n)}(0).\eeq
It follows that, provided the $\beta^{(n)}(0)$ exist, the series will be at worst
asymptotic.

Letting $\lambda_{\R B}^{min}$ be the smallest non-zero eigenvalue and 
using the bound $a^i_{\R B}\le 1$ we find that
\beq{thirteen}\vert\beta^{(n)}(0)\vert <\sum_{\R B\in\C B} z^{N_{\R B}}N_{\R B}
(\lambda_{\R B}^{min})^{-n-2}.\eeq
Now $\lambda_{\R B}^{min}\sim N_{\R B}^{-\delta} $ where the largest value
taken by $\delta$ is  $\delta=2$ (the value for the linear
polymer). This leaves us with an upper bound on $\beta^n(0)$ of the form
\beq{fourteen} \Gamma(2n+ \ts{\frac{9}{2}})(1-4z)^{-2n-9/2}.\eeq
Of course this is a substantial over-estimate compared to the detailed
calculation presented in the body of this paper but it tells us two things.
Firstly the series \rf{seven} is asymptotic. Secondly, since the linear 
polymer is only exponentially suppressed in the ensemble,
 $\beta^n(0)$ must grow much faster
than $\Gamma(n)$ and hence the series cannot be Borel summable.

\end{document}